\documentclass[11pt]{article}
\usepackage{amssymb,amsmath,amsfonts}
\usepackage{graphicx}
\usepackage{graphics}
\usepackage{eepic,epsfig}
\usepackage{dcolumn}

\begin{document}

\title{Synchrotron radiation from a charge moving \\
along a helix around a dielectric cylinder}
\author{A. A. Saharian\thanks{%
E-mail: saharian@ictp.it} \, and A. S. Kotanjyan \\
\textit{Institute of Applied Problems in Physics, 0014 Yerevan, Armenia}}
\date{\today}
\maketitle

\begin{abstract}
In this paper we investigate the radiation emitted by a charged particle
moving along a helical orbit around a dielectric cylinder immersed into a
homogeneous medium. Formulae are derived for the electromagnetic potentials,
electric and magnetic fields, and for the spectral-angular distribution of
the radiation in the exterior medium. It is shown that under the Cherenkov
condition for dielectric permittivity of the cylinder and the velocity of
the particle image on the cylinder surface, strong narrow peaks appear in
the angular distribution for the number of quanta radiated on a given
harmonic. At these peaks the radiated energy exceeds the corresponding
quantity for a homogeneous medium by several dozens of times. Simple
analytic estimates are given for the heights and widths of these peaks. The
results of numerical calculations for the angular distribution of the
radiated quanta are presented and they are compared with the corresponding
quantities for the radiation from a charge moving along a helical trajectory
inside a dielectric cylinder.
\end{abstract}

\bigskip

PACS number(s): 41.60.Ap, 41.60.Bq

\bigskip

\section{Introduction}

\label{sec:oscint}

The extensive applications of synchrotron radiation in a wide variety of
experiments and in many disciplines (see \cite{Koch}-\cite{Hofm04} and
references therein), motivate the importance of investigations for various
mechanisms of controlling the radiation parameters. In particular, it is of
interest to consider the influence of a medium on the spectral and angular
characteristics of the radiation. It is well known that the presence of
medium can essentially change the characteristics of the electromagnetic
processes and gives rise to new types of phenomena such as Cherenkov,
transition, and diffraction radiations. Moreover, the operation of a number
of devices assigned to production of electromagnetic radiation is based on
the interaction of high-energy particles with materials (see, for example,
\cite{Rull98}). The synchrotron radiation from a charged particle
circulating in a homogeneous medium was considered in \cite{Tsytovich},
where it has been shown that the interference between the synchrotron and
Cherenkov radiations leads to remarkable effects. New interesting features
arise in the case of inhomogeneous media.

Interfaces of media are widely used to control the radiation flow emitted by
various systems. Well-known examples of this kind are the Cherenkov
radiation of a charge moving parallel to a plane interface of two media or
flying parallel to the axis of a dielectric cylinder, transition radiation,
Smith-Purcell radiation. In \cite{Mkrt91} the radiation from charged
particles flying over a surface acoustic wave generated on a plane interface
between two media is studied. In a series of papers initiated in \cite%
{Grig95a,Grig95b,Grigoryan1995}, we have investigated the influence of
spherical and cylindrical boundaries between two dielectrics on the
characteristics of the synchrotron radiation. In particular, the
investigation of radiation from a charge rotating about/inside a dielectric
ball showed \cite{Grigoryan1998} that when the Cherenkov condition for the
ball material and particle speed is satisfied, there appear high narrow
peaks in the spectral distribution of the number of quanta emitted to outer
space at some specific values of the ratio of ball-to-particle orbit radii.
In the vicinity of these peaks the rotating particle may generate radiation
field quanta exceeding in several dozens of times those generated by
particle rotating in a continuous, infinite and transparent medium having
the same real part of permittivity as the ball material. Similar problems
with the cylindrical symmetry have been discussed in \cite%
{Grigoryan1995,Kot2000}. It is shown that under similar Cherenkov condition
for permittivity of cylinder and the speed of particle gyrating about/inside
the cylinder, high narrow peaks are present in the spectral-angular
distribution for the number of radiated quanta. In the vicinity of these
peaks the radiated energy exceeds the corresponding value for homogeneous
medium case by several orders of magnitude.

In recent papers \cite{Saha05}-\cite{Arzu07}, we have investigated more
general problem of the radiation from a charged particle in a helical
trajectory inside a dielectric cylinder. The corresponding problem for a
charge moving in vacuum has been widely discussed in the literature (see,
for instance, \cite{Bord99,Hofm04} and references given therein) and the
influence of a homogeneous dispersive medium is considered in \cite{Gevo84}.
This type of electron motion is used in helical undulators for generating
electromagnetic radiation in a narrow spectral interval at frequencies
ranging from radio or millimeter waves to x-rays (see \cite%
{Bord99,Hofm04,Onuk03}).

The present paper is devoted to the investigation of the radiation from a
charge moving along a helix around a dielectric cylinder immersed in a
homogeneous medium. The paper is organized as follows. In the next section,
by using the formulae for the Green function from \cite{Grigoryan1995}, we
derive expressions for the vector potential and electromagnetic fields in
the region outside the cylinder. The angular-frequency distribution of the
radiation intensity in the surrounding medium is considered in section \ref%
{sec:radiation}. In section \ref{sec:Properties} we discuss the features of
the radiation intensity and the results of the numerical evaluations are
presented. Section \ref{sec:Conc} concludes the main results of the paper.

\bigskip

\section{Electromagnetic fields in the exterior region}

\label{sec:oscfields}

Consider a point charge $q$ moving along the helical trajectory of radius $%
\rho _{0}$ outside a dielectric cylinder with radius $\rho _{1}$ and with
permittivity $\varepsilon _{0}$. We will assume that this system is immersed
in a homogeneous medium with dielectric permittivity $\varepsilon _{1}$
(magnetic permeability will be taken to be unit). This type of motion can be
produced by a uniform constant magnetic field directed along the axis of a
cylinder, by a circularly polarized plane wave, or by a spatially periodic
transverse magnetic field of constant absolute value and a direction that
rotates as a function of the longitudinal coordinate. In helical undulators
the last configuration is used. In the cylindrical coordinate system ($\rho
,\phi ,z$) with the axis $z$ directed along the cylinder axis, the
components of the current density created by the charge are given by the
formula
\begin{equation}
j_{l}=\frac{q}{\rho }v_{l}\delta (\rho -\rho _{0})\delta (\phi -\omega
_{0}t)\delta (z-v_{\parallel }t).  \label{hosqxtutjun}
\end{equation}%
where $\omega _{0}=v_{\perp }/\rho _{0}$ is the angular velocity of the
charge, $v_{\parallel }$ and $v_{\perp }$ are the particle velocities along
the axis of the cylinder and in the perpendicular plane, respectively.

The solution to Maxwell equations for the vector potential is expressed in
terms of the Green function $G_{il}(\mathbf{r},t,\mathbf{r}^{\prime
},t^{\prime })$ of the electromagnetic field by the formula
\begin{equation}
A_{i}(\mathbf{r},t)=-\frac{1}{2\pi ^{2}c}\int G_{il}(\mathbf{r},t,\mathbf{r}%
^{\prime },t^{\prime })j_{l}(\mathbf{r}^{\prime },t^{\prime })d\mathbf{r}%
^{\prime }dt^{\prime },  \label{vecpot}
\end{equation}%
where the summation over $l$ is understood. In accordance with the problem
symmetry, for the Green function we have the following Fourier expansion:
\begin{eqnarray}
G_{il}(\mathbf{r},t,\mathbf{r}^{\prime },t^{\prime })
&=&\sum_{m=-\infty }^{\infty }\int_{-\infty }^{\infty
}dk_{z}d\omega G_{il}(m,k_{z},\omega
,\rho ,\rho ^{\prime })  \notag \\
&&\times \exp [im(\phi -\phi ^{\prime })+ik_{z}(z-z^{\prime })-i\omega
(t-t^{\prime })].  \label{GF_furieexp}
\end{eqnarray}%
Substituting expressions (\ref{hosqxtutjun}) and (\ref{GF_furieexp}) into
formula (\ref{vecpot}), we present the vector potential as the Fourier
expansion
\begin{equation}
A_{l}(\mathbf{r},t)=\sum_{m=-\infty }^{\infty }e^{im(\phi -\omega
_{0}t)}\int_{-\infty }^{\infty }dk_{z}e^{ik_{z}(z-v_{\parallel
}t)}A_{ml}(m,k_{z},\rho ),  \label{vecpot2}
\end{equation}%
where the coefficients $A_{ml}=A_{ml}(m,k_{z},\rho )$ are given in terms of
the Fourier components of the Green function by the formula
\begin{equation}
A_{ml}=-\frac{q}{\pi c}\left[ v_{\perp }G_{l\phi }(m,k_{z},\omega
_{m}(k_{z}),\rho ,\rho _{0})+v_{\parallel }G_{lz}(m,k_{z},\omega
_{m}(k_{z}),\rho ,\rho _{0})\right] .  \label{vecpot1new}
\end{equation}%
Here and in the discussion below we use the notation
\begin{equation}
\omega _{m}(k_{z})=m\omega _{0}+k_{z}v_{\parallel }.  \label{omegam}
\end{equation}

By making use of the formula for the Green function given earlier in \cite%
{Grigoryan1995}, in the Lorentz gauge for the corresponding Fourier
components $G_{il}=G_{il}(m,k_{z},\omega _{m}(k_{z}),\rho ,\rho _{0})$ in
the region outside the cylinder, $\rho >\rho _{1}$, we find
\begin{eqnarray}
G_{l\phi } &=&\frac{i^{2-\sigma _{l}}}{2}\sum_{p=\pm 1}p^{\sigma
_{l}}B_{m}^{(p)}H_{m+p}(\lambda _{1}\rho ),  \notag \\
G_{lz} &=&\frac{i^{2-\sigma _{l}}k_{z}}{2\rho _{1}}\frac{H_{m}(\lambda
_{1}\rho _{0})J_{m}(\lambda _{0}\rho _{1})}{\alpha _{m}V_{m}^{H}}\sum_{p=\pm
1}p^{\sigma _{l}-1}\frac{J_{m+p}(\lambda _{0}\rho _{1})}{V_{m+p}^{H}}%
H_{m+p}(\lambda _{1}\rho ),  \label{GFcompd} \\
G_{zz} &=&\frac{\pi }{2iV_{m}^{H}}\left[ J_{m}(\lambda _{1}\rho
_{0})V_{m}^{H}-H_{m}(\lambda _{1}\rho _{0})V_{m}^{J}\right] H_{m}(\lambda
_{1}\rho ),  \notag
\end{eqnarray}%
where in the first and second equations $l=\rho ,\phi $, $\sigma _{\rho }=1$%
, $\sigma _{\phi }=2$,
\begin{equation}
\lambda _{j}^{2}=\frac{\omega _{m}^{2}(k_{z})}{c^{2}}\varepsilon
_{j}-k_{z}^{2},\quad j=0,1,  \label{lambdaj}
\end{equation}%
$J_{m}(x)$ is the Bessel function, $H_{m}(x)=H_{m}^{(1)}(x)$ is the Hankel
function of the first kind, and
\begin{equation}
V_{m}^{F}=J_{m}(\lambda _{0}\rho _{1})\frac{\partial F_{m}(\lambda _{1}\rho
_{1})}{\partial \rho _{1}}-F_{m}(\lambda _{1}\rho _{1})\frac{\partial
J_{m}(\lambda _{0}\rho _{1})}{\partial \rho _{1}},\;F=J,H.  \label{Wronsk}
\end{equation}%
In formulae (\ref{GFcompd}) the coefficients $B_{m}^{(p)}$, $p=\pm 1$, are
defined by the expressions
\begin{eqnarray}
B_{m}^{(p)} &=&\frac{\pi }{2iV_{m+p}^{H}}\left[ J_{m+p}(\lambda _{1}\rho
_{0})V_{m+p}^{H}-H_{m+p}(\lambda _{1}\rho _{0})V_{m+p}^{J}\right]  \notag \\
&&+\frac{p\lambda _{0}J_{m+p}(\lambda _{0}\rho _{1})J_{m}(\lambda _{0}\rho
_{1})}{2\rho _{1}\alpha _{m}V_{m+p}^{H}}\sum_{l=\pm 1}\frac{H_{m+l}(\lambda
_{1}\rho _{0})}{V_{m+l}^{H}},  \label{Bm+p}
\end{eqnarray}%
with the notation
\begin{equation}
\alpha _{m}=\frac{\varepsilon _{0}}{\varepsilon _{1}-\varepsilon _{0}}-\frac{%
\lambda _{0}J_{m}(\lambda _{0}\rho _{1})}{2}\sum_{l=\pm 1}l\frac{%
H_{m+l}(\lambda _{1}\rho _{1})}{V_{m+l}^{H}}.  \label{bet1}
\end{equation}%
Note that the equation $\alpha _{m}=0$ determines the eigenmodes of the
dielectric cylinder.

In the definition of $\lambda _{1}$ given by (\ref{lambdaj}), one should
take into account that in the presence of the imaginary part $\varepsilon
_{1}^{\prime \prime }(\omega )$ for the dielectric permittivity ($%
\varepsilon _{1}=\varepsilon _{1}^{\prime }+i\varepsilon _{1}^{\prime \prime
}$) the radiation field in the exterior medium should exponentially damp for
large $\rho $. This leads to the following relations:
\begin{equation}
\lambda _{1}=%
\begin{cases}
(\omega _{m}/c)\sqrt{\varepsilon _{1}-k_{z}^{2}c^{2}/\omega
_{m}^{2}}, &
\omega _{m}^{2}\varepsilon _{1}/c^{2}>k_{z}^{2}, \\
i\sqrt{k_{z}^{2}-\omega _{m}^{2}\varepsilon _{1}/c^{2}}, & \omega
_{m}^{2}\varepsilon _{1}/c^{2}<k_{z}^{2}.%
\end{cases}
\label{lambda1}
\end{equation}%
For $\lambda _{1}^{2}>0$ the sign of $\lambda _{1}$ may also be determined
from the principle of radiation (different signs of $\omega t$ and $\lambda
_{1}\rho $ in the expressions of the fields) for large $\rho $.

By taking into account formulae (\ref{GFcompd}), from (\ref{vecpot1new}) we
obtain the expressions for the Fourier components of the vector potential:
\begin{eqnarray}
A_{ml} &=&-\frac{qi^{2-\sigma _{l}}}{2\pi }\sum_{p=\pm 1}p^{\sigma
_{l}}C_{m}^{(p)}H_{m+p}(\lambda _{1}\rho ),\;l=\rho ,\phi ,  \notag \\
A_{mz} &=&-\frac{qv_{\parallel }}{2iV_{m}^{H}}\left[ J_{m}(\lambda _{1}\rho
_{0})V_{m}^{H}-H_{m}(\lambda _{1}\rho _{0})V_{m}^{J}\right] H_{m}(\lambda
_{1}\rho ),  \label{vecpot3c}
\end{eqnarray}%
with the coefficients
\begin{equation}
C_{m}^{(p)}=\frac{v_{\perp }}{c}B_{m}^{(p)}+pv_{\parallel }k_{z}\frac{%
J_{m}(\lambda _{0}\rho _{1})J_{m+p}(\lambda _{0}\rho _{1})H_{m}(\lambda
_{1}\rho _{0})}{c\rho {_{1}}\alpha _{m}V_{m}^{H}V_{m+p}^{H}}.  \label{Cm+p}
\end{equation}

The Fourier expansions similar to (\ref{vecpot2}) may also be written for
the electric and magnetic fields. In the case of the magnetic field the
corresponding Fourier coefficients have the form
\begin{eqnarray}
H_{ml} &=&\frac{i^{2-\sigma _{l}}qk_{z}}{2\pi }\sum_{p=\pm 1}p^{\sigma
_{l}-1}D_{m}^{(p)}H_{m+p}(\lambda _{1}\rho ),\;l=\rho ,\phi ,  \notag \\
H_{mz} &=&-\frac{q\lambda _{1}}{2\pi }\sum_{p=\pm
1}pD_{m}^{(p)}H_{m}(\lambda _{1}\rho ),  \label{magneticc}
\end{eqnarray}%
with the notation
\begin{equation}
D_{m}^{(p)}=C_{m}^{(p)}-\frac{v_{\parallel }\lambda _{1}\pi }{%
2ick_{z}V_{m}^{H}}\left[ J_{m}(\lambda _{1}\rho _{0})V_{m}^{H}-H_{m}(\lambda
_{1}\rho _{0})V_{m}^{J}\right] ,\quad p=\pm 1.  \label{Dm}
\end{equation}%
The corresponding Fourier coefficients for the electric field are obtained
from the Maxwell equation $\nabla \times \mathbf{H}=-i\omega \varepsilon _{1}%
\mathbf{E}/c$ with the result:
\begin{eqnarray}
E_{ml} &=&\frac{i^{1-\sigma _{l}}qc}{4\pi \omega _{m}(k_{z})\varepsilon _{1}}%
\sum_{p=\pm 1}p^{\sigma _{l}}H_{m+p}(\lambda _{1}\rho )\left\{ \left( \frac{%
\omega _{m}^{2}(k_{z})\varepsilon _{1}}{c^{2}}+k_{z}^{2}\right)
D_{m}^{(p)}-\lambda
_{1}^{2}D_{m}^{(-p)}\right\} ,  \notag \\
E_{mz} &=&\frac{icq\lambda _{1}k_{z}}{2\pi \omega _{m}(k_{z})\varepsilon _{1}%
}\sum_{p=\pm 1}D_{m}^{(p)}H_{m}(\lambda _{1}\rho ),  \label{electricc}
\end{eqnarray}%
where, as before, $l=\rho ,\phi $. Note that for purely transversal motion ($%
v_{\parallel }=0$) one has $D_{m}^{(p)}=v_{\perp }B_{m}^{(p)}/c$. For a
particle motion in a homogeneous medium we have $\varepsilon
_{0}=\varepsilon _{1}$ and the terms in the expressions of the coefficients $%
C_{m}^{(p)}$ and $B_{m}^{(p)}$ involving the function $\alpha _{m}$ vanish.
In this case $V_{m}^{J}=0$, $V_{m}^{H}=2i/\pi \rho _{1}$, and for the
coefficients $D_{m}^{(p)}$ one finds
\begin{equation}
D_{m}^{(p)}=\frac{\pi }{2i}\left[ \frac{v_{\perp }}{c}J_{m+p}(\lambda
_{1}\rho _{0})-\frac{v_{\parallel }\lambda _{1}}{ck_{z}}J_{m}(\lambda
_{1}\rho _{0})\right] ,\quad \varepsilon _{0}=\varepsilon _{1}.
\label{Dmpham}
\end{equation}

The Fourier coefficients for the fields determined by relations (\ref%
{magneticc}), (\ref{electricc}) have poles corresponding to the zeros of the
function $\alpha _{m}$ appearing in the denominators of (\ref{Bm+p}) and (%
\ref{Cm+p}). It can be seen that this function has zeros only under the
conditions $\lambda _{1}^{2}<0<\lambda _{0}^{2}$. In particular, from here,
as a necessary condition, we should have $\varepsilon _{1}<\varepsilon _{0}$%
. For the corresponding modes the Fourier coefficients of the fields depend
on the radial coordinate through the MacDonald function $K_{\nu }(|\lambda
_{1}|\rho )$ with $\nu =m,m\pm 1$, and they are exponentially damped with
the distance from the cylinder surface. These modes are the eigenmodes of
the dielectric cylinder and they correspond to the waves propagating inside
the cylinder. Below, in the consideration of the intensity for the radiation
to the exterior medium, we will neglect the contribution of the poles
corresponding to these modes.

\section{Radiation intensity}

\label{sec:radiation}

Having the electromagnetic fields we can investigate the intensity of the
radiation propagating in the exterior medium. As we have mentioned before,
for $\lambda _{1}^{2}<0$ the corresponding Fourier components are
exponentially damped for large values $\rho $, and the radiation is present
only under the condition $\lambda _{1}^{2}>0$. The average energy flux per
unit time through the cylindrical surface of radius $\rho $, coaxial with
the dielectric cylinder, is given by the Poynting vector:
\begin{equation}
I=\frac{c}{2T}\int_{0}^{T}dt\int_{-\infty }^{\infty }\left[ \mathbf{E}\times
\mathbf{H}\right] \cdot \mathbf{n}_{\rho }\rho dz,  \label{Poyntingvector}
\end{equation}%
being $T=2\pi /\omega _{0}$ the period for the transverse motion of the
charge. Substituting the corresponding Fourier expansions of the fields and
using the asymptotic expressions of the Hankel functions, at large distances
from the cylinder for the radiation intensity we find
\begin{equation}
I=\frac{q^{2}c^{2}}{\pi ^{2}}\sideset{}{'}{\sum}_{m=0}^{\infty
}\int_{\lambda _{1}^{2}>0}\frac{dk_{z}}{\varepsilon _{1}|\omega _{m}(k_{z})|}%
\left[ \frac{\omega _{m}^{2}(k_{z})}{c^{2}}\varepsilon _{1}\left\vert
D_{m}^{(+1)}-D_{m}^{(-1)}\right\vert ^{2}+k_{z}^{2}\left\vert
D_{m}^{(+1)}+D_{m}^{(-1)}\right\vert ^{2}\right] ,  \label{Int10}
\end{equation}%
where the prime over the sum means that the term with $m=0$ should be taken
with the weight $1/2$ and the coefficients $D_{m}^{(p)}$ are defined by
formulae (\ref{Dm}).

First let us consider the special case $\omega _{0}=0$ for a fixed $\rho
_{0} $, which corresponds to a charge moving with constant velocity $%
v_{\parallel }$ on a straight line $\rho =\rho _{0}$ parallel to the
cylinder axis. For this case, from (\ref{omegam}), $\omega
_{m}(k_{z})=k_{z}v_{\parallel }$ and expressions (\ref{lambdaj}) take the
form $\lambda _{j}^{2}=k_{z}^{2}(\beta _{j\parallel }^{2}-1)$ with the
notation%
\begin{equation}
\beta _{j\parallel }=\frac{v_{\parallel }}{c}\sqrt{\varepsilon _{j}},\;j=0,1.
\label{betj}
\end{equation}%
From the condition $\lambda _{1}^{2}>0$ it follows that the radiation in the
exterior medium is present only under the Cherenkov condition for the
particle velocity, $v_{\parallel }$, and dielectric permittivity $%
\varepsilon _{1}$ for the surrounding medium: $\beta _{1\parallel }>1$.
Introducing the angle $\vartheta $ of the wave vector with the cylinder
axis, from the relation $k_{z}=\omega /v_{\parallel }$ it follows that $\cos
\vartheta =\beta _{1\parallel }^{-1}$, and the radiation propagates along
the Cherenkov cone of the external medium. Since for the case under
consideration one has $v_{\perp }=0$, the first term on the right of formula
(\ref{Cm+p}) vanishes and the expression for the coefficients $D_{m}^{(p)}$
is written in the form%
\begin{eqnarray}
D_{m}^{(p)} &=&-\frac{v_{\parallel }\pi }{2ic}\sqrt{\beta _{1\parallel
}^{2}-1}J_{m}(\lambda _{1}^{(0)}\rho _{0})+v_{\parallel }\frac{H_{m}(\lambda
_{1}^{(0)}\rho _{0})}{cV_{m}^{H}}  \notag \\
&&\times \left[ pk_{z}\frac{J_{m}(\lambda _{0}^{(0)}\rho
_{1})J_{m+p}(\lambda _{0}^{(0)}\rho _{1})}{\rho {_{1}}\alpha _{m}V_{m+p}^{H}}%
+\frac{\pi }{2i}V_{m}^{J}\sqrt{\beta _{1\parallel }^{2}-1}\right] ,
\label{Dmpm0}
\end{eqnarray}%
where
\begin{equation}
\lambda _{1}^{(0)}=k_{z}\sqrt{\beta _{1\parallel }^{2}-1},\;\lambda
_{0}^{(0)}=%
\begin{cases}
k_{z}\sqrt{\beta _{0\parallel }^{2}-1}, & \beta _{0\parallel }>1, \\
i|k_{z}|\sqrt{1-\beta _{0\parallel }^{2}}, & \beta _{0\parallel }<1.%
\end{cases}
\label{lambj}
\end{equation}%
The substitutions $\lambda _{j}\rightarrow \lambda _{j}^{(0)}$ should also
be made in the formulae for $V_{m}^{J}$, $V_{m+p}^{H}$, and $\alpha _{m}$.
Changing the integration variable, from (\ref{Int10}) for the corresponding
radiation intensity we find

\begin{equation}
I=\frac{2q^{2}}{\pi ^{2}v_{\parallel }}\sideset{}{'}{\sum}_{m=0}^{\infty
}\int_{\beta _{1\parallel }>1}d\omega \,\omega \left[ \left\vert
D_{m}^{(+1)}-D_{m}^{(-1)}\right\vert ^{2}+\beta _{1\parallel
}^{-2}\left\vert D_{m}^{(+1)}+D_{m}^{(-1)}\right\vert ^{2}\right] ,
\label{IOm0}
\end{equation}%
where in formula (\ref{Dmpm0}) for the coefficients $D_{m}^{(p)}$ the
expressions (\ref{lambj}) should be used with $k_{z}=\omega /v_{\parallel }$%
. In the limit $\rho _{1}\rightarrow 0$ the only contribution to the
radiation intensity comes from the first term on the right-hand side of
formula (\ref{Dmpm0}) and, after the summation over $m$ by using the formula
$\sum _{m=-\infty }^{+\infty }J_{m}^{2}(x)=1$, we obtain the standard
expression for the intensity of the Cherenkov radiation in a homogeneous
medium. For small values of the cylinder radius, the part in the radiation
intensity induced by the presence of the cylinder vanishes as $\rho
_{1}^{2m} $ for the harmonics $m\geqslant 1$ and as $\rho _{1}^{2}$ for $m=0$%
.

Now we return to the general case of the particle motion along the helix.
First we consider the contribution of the mode with $m=0$ to the radiation
intensity given by (\ref{Int10}). For this mode one has $\omega
_{m}(k_{z})=k_{z}v_{\parallel }$ and, as in the previous case, from the
condition $\lambda _{1}^{2}>0$ it follows that in the exterior region, the
corresponding radiation is present only under the condition $\beta
_{1\parallel }>1$. This radiation propagates along the Cherenkov cone $%
\vartheta =\vartheta _{0}\equiv \arccos (\beta _{1\parallel }^{-1})$ of the
external medium. By using the expressions for the coefficients $D_{m}^{(p)}$
and introducing a new integration variable $\omega =|k_{z}|v_{\parallel }$,
for the radiation intensity at $m=0$ one obtains%
\begin{equation}
I_{m=0}=\frac{q^{2}}{v_{\parallel }}\int_{\beta _{1\parallel }>1}\frac{%
\omega d\omega }{\varepsilon _{1}}\bigg[\beta _{1\perp }^{2}\left\vert
U_{1}(\omega )\right\vert ^{2}+\left\vert U_{2}(\omega )+i\sqrt{\beta
_{1\parallel }^{2}-1}U_{0}(\omega )\right\vert ^{2}\bigg],  \label{Intm0}
\end{equation}%
with the notations%
\begin{eqnarray}
U_{l}(\omega ) &=&J_{l}(\lambda _{1}^{(0)}\rho _{0})-H_{l}(\lambda
_{1}^{(0)}\rho _{0})\frac{V_{l}^{J}}{V_{l}^{H}},\;l=0,1,  \notag \\
U_{2}(\omega ) &=&\frac{2(\varepsilon _{0}-\varepsilon _{1})\omega
J_{0}(\lambda _{0}^{(0)}\rho _{1})J_{1}(\lambda _{0}^{(0)}\rho
_{1})H_{0}(\lambda _{1}^{(0)}\rho _{0})(\pi \rho {_{1}}V_{0}^{H})^{-1}}{%
\varepsilon _{1}\lambda _{0}^{(0)}J_{0}(\lambda _{0}^{(0)}\rho
_{1})H_{1}(\lambda _{1}^{(0)}\rho _{1})-\varepsilon _{0}\lambda
_{1}J_{1}(\lambda _{0}^{(0)}\rho _{1})H_{0}(\lambda _{1}^{(0)}\rho _{1})}.
\label{Ulom}
\end{eqnarray}%
In formulae (\ref{Ulom}), $\lambda _{j}^{(0)}$ are defined by relations (\ref%
{lambj}) with $k_{z}=\omega /v_{\parallel }$ and
\begin{equation}
\beta _{j\perp }=\frac{v_{\perp }}{c}\sqrt{\varepsilon _{j}}.
\label{betjperp}
\end{equation}

In the limit $\rho _{1}\rightarrow 0$, by using the formulae for the Bessel
functions for small values of the argument, from formula (\ref{Intm0}) we
find
\begin{equation}
I_{m=0}\approx I_{0}^{(0)}+\frac{\pi q^{2}\rho _{1}^{2}}{2v_{\parallel }}%
\int_{\beta _{1\parallel }>1}d\omega \frac{\varepsilon _{1}-\varepsilon _{0}%
}{\varepsilon _{1}^{2}}\omega ^{3}\left( \beta _{1\parallel }^{2}-1\right)
^{2}J_{0}(\lambda _{1}\rho _{0})Y_{0}(\lambda _{1}\rho _{0}),
\label{Intm0Lim}
\end{equation}%
where $Y_{0}(x)$ is the Neumann function, and%
\begin{equation}
I_{0}^{(0)}=\frac{q^{2}}{v_{\parallel }}\int_{\beta _{1\parallel }>1}d\omega
\frac{\omega }{\varepsilon _{1}}\left[ \beta _{1\perp }^{2}J_{1}^{2}\left(
\lambda _{1}^{(0)}\rho _{0}\right) +(\beta _{1\parallel
}^{2}-1)J_{0}^{2}\left( \lambda _{1}^{(0)}\rho _{0}\right) \right] ,
\label{Intm0ham}
\end{equation}%
is the radiation intensity at $m=0$ in the case of the charge motion in a
homogeneous medium with $\varepsilon _{0}=\varepsilon _{1}$.

For the radiation at $m\neq 0$ harmonics, from the condition $\lambda
_{1}^{2}>0$ one obtains the quadratic inequality with respect to $k_{z}$:
\begin{equation}
k_{z}^{2}(1-\beta _{1\parallel }^{-2})+2k_{z}m\omega _{0}/v_{\parallel
}+(m\omega _{0}/v_{\parallel })^{2}>0.  \label{qarakusihav}
\end{equation}%
It is convenient to write the solution to this inequality in terms of a new
variable $\vartheta $, $0\leqslant \vartheta \leqslant \pi $, defined in
accordance with the relation
\begin{equation}
k_{z}=\frac{m\omega _{0}}{c}\frac{\sqrt{\varepsilon _{1}}\cos \vartheta }{%
1-\beta _{1\parallel }\cos \vartheta }.  \label{kazet}
\end{equation}%
Now the function $\omega _{m}(k_{z})$ is presented in the form
\begin{equation}
\omega _{m}(k_{z})=\frac{m\omega _{0}}{1-\beta _{1\parallel }\cos \vartheta }%
,  \label{omegatet}
\end{equation}%
and the quantities $k_{z}$ and $\omega _{m}(k_{z})$ are connected by the
relation $k_{z}=\omega _{m}(k_{z})\sqrt{\varepsilon _{1}}\cos \vartheta \,/c$%
. In terms of the new variable $\vartheta $, at large distances from the
charge trajectory the dependence of elementary waves on the space-time
coordinates has the form%
\begin{equation}
\exp [\omega _{m}(k_{z})\sqrt{\varepsilon _{1}}(\rho \sin \vartheta +z\cos
\vartheta -ct/\sqrt{\varepsilon _{1}})/c],  \label{timedep}
\end{equation}%
which describes wave with the frequency
\begin{equation}
\omega _{m}=\left\vert \omega _{m}(k_{z})\right\vert =\frac{m\omega _{0}}{%
|1-\beta {_{1{\parallel }}}\cos \vartheta |},\;m=1,2,\ldots ,
\label{omegamtet}
\end{equation}%
propagating at the angle $\vartheta $ to the $z$-axis. Formula (\ref%
{omegamtet}) describes the normal Doppler effect in the cases $\beta
_{1\parallel }<1$ and $\beta _{1\parallel }>1$, $\vartheta >\vartheta _{0}$,
and anomalous Doppler effect inside the Cherenkov cone, $\vartheta
<\vartheta _{0}$, in the case $\beta _{1\parallel }>1$.

Changing the integration over $k_{z}$ in (\ref{Int10}) to the integration
over $\vartheta $ in accordance with (\ref{kazet}), the radiation intensity
at $m\neq 0$ harmonics is presented in the form
\begin{equation}
I_{m\neq 0}=\sum_{m=1}^{\infty }\int d\Omega \,\frac{dI_{m}}{d\Omega },
\label{Int3a}
\end{equation}%
where $d\Omega =\sin \vartheta d\vartheta d\phi $ is the solid angle
element, and
\begin{equation}
\frac{dI_{m}}{d\Omega }=\frac{q^{2}\omega _{0}^{2}m^{2}\sqrt{\varepsilon _{1}%
}}{2\pi ^{3}c|1-\beta _{1\parallel }\cos \vartheta |^{3}}\left[ \left\vert
D_{m}^{(1)}-D_{m}^{(-1)}\right\vert ^{2}+\left\vert
D_{m}^{(1)}+D_{m}^{(-1)}\right\vert ^{2}\cos ^{2}\vartheta \right] ,
\label{Int3}
\end{equation}%
is the average power radiated by the charge at a given harmonic $m$ into a
unit solid angle. In formulae (\ref{Dm}) for the coefficients $D_{m}^{(\pm
1)}$, the quantities $\lambda _{0}$ and $\lambda _{1}$ are expressed in
terms of $\vartheta $ as
\begin{eqnarray}
&&\lambda _{0}=\frac{m\omega _{0}}{c}\frac{\sqrt{\varepsilon
_{0}-\varepsilon _{1}\cos ^{2}\vartheta }}{1-\beta _{1\parallel }\cos
\vartheta },  \notag \\
&&\lambda _{1}=\frac{m\omega _{0}}{c}\frac{\sqrt{\varepsilon _{1}}\sin
\vartheta }{1-\beta _{1\parallel }\cos \vartheta }.  \label{lambondb}
\end{eqnarray}

Hence, under the Cherenkov condition, $\beta _{1\parallel }>1$, the total
radiation intensity is presented in the form
\begin{equation}
I=I_{0}+I_{m\neq 0}\,,  \label{totintOm}
\end{equation}%
where the first term on the right-hand side is given by formula (\ref{Intm0}%
) and describes the radiation with a continuous spectrum propagating along
the Cherenkov cone $\vartheta =\vartheta _{0}$ of the external medium. The
second term describes the radiation, which, for a given angle $\vartheta $,
has discrete spectrum determined by formula (\ref{omegamtet}). If the
Cherenkov condition is not satisfied ($\beta _{1\parallel }<1$) the first
term is absent and only the modes with $m\neq 0$ contribute to the radiation
intensity.

For a charge moving in a homogeneous medium with dielectric permittivity $%
\varepsilon _{1}$, one has $\varepsilon _{0}=\varepsilon _{1}$, and using
formula (\ref{Dmpham}) for the coefficients $D_{m}^{(p)}$, from formula (\ref%
{Int3}) we obtain
\begin{equation}
\frac{dI_{m}^{(0)}}{d\Omega }=\frac{q^{2}\omega _{0}^{2}m^{2}}{2\pi c\sqrt{%
\varepsilon _{1}}|1-\beta _{1\parallel }\cos \vartheta |^{3}}\left[ \beta
_{1\perp }^{2}J_{m}^{\prime }{}^{2}(\lambda _{1}\rho _{0})+\left( \frac{\cos
\vartheta -\beta _{1\parallel }}{\sin \vartheta }\right)
^{2}J_{m}^{2}(\lambda _{1}\rho _{0})\right] .  \label{Intham}
\end{equation}%
In the case $\varepsilon _{1}=1$ this formula is derived in Ref. \cite%
{Soko68} (see also Refs. \cite{Soko86,Tern85,Bord99}). For a charge with a
purely transversal motion ($v_{\parallel }=0$) one has $D_{m}^{(p)}=(v_{%
\perp }/c)B_{m}^{(p)}$, and from (\ref{Int3}) as a special case we obtain
the formula derived in \cite{Kot2000}.

\section{Features of the radiation}

\label{sec:Properties}

In this section, on the base of the formulae given before, we discuss the
characteristic features of the radiation intensity. For a non-relativistic
motion, $\beta _{1\perp },\beta _{1\parallel }\ll 1$, from the general
formula (\ref{Int3}) we find
\begin{equation}
\frac{dI_{m}}{d\Omega }\approx \frac{2q^{2}c(m\beta _{1\perp }/2)^{2(m+1)}}{%
\pi \rho _{0}^{2}\varepsilon _{1}^{3/2}(m!)^{2}}\left[ 1+\frac{\varepsilon
_{1}-\varepsilon _{0}}{\varepsilon _{0}+\varepsilon _{1}}\left( \frac{\rho
_{1}}{\rho _{0}}\right) ^{2m}\right] ^{2}(1+\cos ^{2}\vartheta )\sin
^{2(m-1)}\vartheta ,  \label{Intnonrel}
\end{equation}%
and the contribution of the harmonics with $m>1$ is small compared to that
in the fundamental one, $m=1$. Note that the part of the radiation intensity
with the first term in square brackets of (\ref{Intnonrel}) corresponds to
the case of the motion in homogeneous medium with permittivity $\varepsilon
_{1}$, and the part with the second term is induced by the presence of the
cylinder with permittivity $\varepsilon _{0}$.

In the limit $\rho _{1}\rightarrow 0$, in the general formula (\ref{Int3})
we use the asymptotic expressions for the Bessel functions for small values
of the argument. After long calculations it can be seen that the difference
between the radiation intensities in the cases when the cylinder is present
and absent, $dI_{m}/d\Omega -dI_{m}^{(0)}/d\Omega $, vanishes as $\rho
_{1}^{2m}$ for $m\geqslant 1$. In the same limit and for the radiation
corresponding to $m=0$, the part induced by the cylinder is given in formula
(\ref{Intm0Lim}) and vanishes like $\rho _{1}^{2}$.

Now let us consider the behavior of the radiation intensity near the
Cherenkov angle when $|1-\beta _{1\parallel }\cos \vartheta |\ll 1$. Using
the asymptotic formulae for the cylinder functions for large values of the
argument it can be seen that in this limit
\begin{equation}
\frac{dI_{m}}{d\Omega }\propto |1-\beta _{1\parallel }\cos \vartheta |^{-2}.
\label{dImasnearCh1}
\end{equation}%
Note that for the radiation intensity in a homogeneous medium with
dielectric permittivity $\varepsilon _{1}$ we have the same behavior. In
accordance with (\ref{omegamtet}), near the Cherenkov cone the frequencies
of the radiated photons are large and the dispersion of the dielectric
permittivity $\varepsilon _{1}$ should be taken into account. The
corresponding angles are determined implicitly from the condition $\omega
_{m}\gtrsim \omega _{d}$, by using formula (\ref{omegamtet}) and frequency
dependence of the permittivity $\varepsilon _{1}=\varepsilon _{1}(\omega
_{m})$. Here $\omega _{d}$ is the characteristic frequency of the
dispersion. Note that for the charge helical motion inside the dielectric
cylinder ($\rho _{0}<\rho _{1}$) the behavior of the radiation intensity
near the Cherenkov cone is radically different for the cases $\beta
_{0\parallel }>1$ and $\beta _{0\parallel }<1$ (see \cite{Saha05}). In the
first case the intensity behaves like $|1-\beta _{1\parallel }\cos \vartheta
|^{-4}$, whereas in the second case the intensity behaves as $|1-\beta
_{1\parallel }\cos \vartheta |^{-4}\exp [-2(\omega _{m}/v_{\parallel })(\rho
_{1}-\rho _{0})\sqrt{1-\beta _{0\parallel }^{2}}]$, with $\omega _{m}$ given
by (\ref{omegamtet}).

By using Debye's asymptotic expansions for the Bessel and Neumann functions,
in \cite{Saha05}, in the case of helical motion inside a dielectric
cylinder, it has been shown that under the condition $|\lambda _{1}|\rho
_{1}<m$, at points where the real part of the function $\alpha _{m}$, given
by formula (\ref{bet1}), is equal to zero, the contribution of the imaginary
part of this function into the coefficients $D_{m}^{(p)}$ can be
exponentially large for large values $m$. This leads to the appearance of
strong narrow peaks in the angular distribution for the radiation intensity
at a given harmonic $m$. The condition for the real part of the function $%
\alpha _{m}$ to be zero has the form:%
\begin{equation}
\sum_{l=\pm 1}\left[ \frac{\lambda _{1}}{\lambda _{0}}\frac{J_{m+l}(\lambda
_{0}\rho _{1})Y_{m}(\lambda _{1}\rho _{1})}{J_{m}(\lambda _{0}\rho
_{1})Y_{m+l}(\lambda _{1}\rho _{1})}-1\right] ^{-1}=\frac{2\varepsilon _{0}}{%
\varepsilon _{1}-\varepsilon _{0}},  \label{peakscondn}
\end{equation}%
where $Y_{\nu }(x)$ is the Neumann function. This equation is obtained from
the equation determining the eigenmodes for the dielectric cylinder by the
replacement $H_{m}\rightarrow Y_{m}$. Equation (\ref{peakscondn}) has no
solutions for $\lambda _{0}^{2}<0$, which is possible only for $\varepsilon
_{0}<\varepsilon _{1}$. Hence, the above mentioned possibility for the
appearance of peaks is not realized for the case $\lambda _{0}^{2}<0$ which,
in accordance with (\ref{lambondb}), corresponds to the angular region $\cos
^{2}\vartheta >\varepsilon _{0}/\varepsilon _{1}$.

Under the condition (\ref{peakscondn}) for the coefficient $\alpha _{m}$ one
has%
\begin{equation}
\alpha _{m}\approx \frac{i\lambda _{0}J_{m}(\lambda _{0}\rho _{1})}{\pi \rho
_{1}}\sum_{l=\pm 1}l\frac{J_{m+l}(\lambda _{0}\rho _{1})}{(V_{m+l}^{Y})^{2}},
\label{alphapeaks}
\end{equation}%
where $V_{m}^{Y}$ is defined by formula (\ref{Wronsk}) with $F=Y$. For large
values $m$, from (\ref{alphapeaks}) we have the estimate $\alpha _{m}\propto
\exp \left[ -2m\zeta (\lambda _{1}\rho _{1}/m)\right] $, with%
\begin{equation}
\zeta (z)=\ln \frac{1+\sqrt{1-z^{2}}}{z}-\sqrt{1-z^{2}},  \label{zetaz}
\end{equation}%
and this coefficient is exponentially small. Now, by using the asymptotic
formulae for the Bessel functions, we can see that under the conditions%
\begin{equation}
|\lambda _{1}|\rho _{0}<m<\lambda _{0}\rho _{1},  \label{peakscondn2}
\end{equation}%
for the angles being the solutions of the equation (\ref{peakscondn}), one
has $D_{m}^{(p)}\propto \exp \left[ m\zeta (\lambda _{1}\rho _{0}/m)\right] $
and, hence, for the radiation intensity%
\begin{equation}
\frac{dI_{m}}{d\Omega }\propto \exp \left[ 2m\zeta (\lambda _{1}\rho _{0}/m)%
\right] .  \label{Npeak}
\end{equation}

Note the when the charge moves inside the cylinder ($\rho _{0}<\rho _{1}$),
for the appearance of the peaks we have two possibilities \cite{Saha05}. In
the first case, corresponding to $|\lambda _{1}|\rho _{1}<m<|\lambda
_{0}|\rho _{0}$, at the peaks the radiation intensity behaves as $\exp \left[
2m\zeta (\lambda _{1}\rho _{1}/m)\right] $. By taking into account that the
function $\zeta (z)$ is monotonically decreasing and comparing this estimate
with (\ref{Npeak}), we conclude that in this case the peaks in the radiation
intensity for the motion inside the cylinder are stronger. The second case
corresponds to $|\lambda _{1}|\rho _{1},|\lambda _{0}|\rho _{0}<m<|\lambda
_{0}|\rho _{1}$, and the radiation intensity at the peaks behaves like $\exp
\{2m[\zeta (\lambda _{1}\rho _{1}/m)-\zeta (\lambda _{0}\rho _{0}/m)]\}$.

From (\ref{peakscondn2}), by taking into account formulae (\ref{lambondb}),
as necessary conditions for the presence of the strong narrow peaks in the
angular distribution for the radiation intensity one has
\begin{equation}
\frac{\omega _{0}\rho _{0}}{c}\sqrt{\varepsilon _{1}}\sin \vartheta
<|1-\beta _{1\parallel }\cos \vartheta |<\frac{\omega _{0}\rho _{1}}{c}\sqrt{%
\varepsilon _{0}-\varepsilon _{1}\cos ^{2}\vartheta }.  \label{peakscond3}
\end{equation}%
These conditions can be satisfied only if we have
\begin{equation}
\varepsilon _{0}>\varepsilon _{1},\quad \tilde{v}\sqrt{\varepsilon _{0}}/c>1,
\label{peakscond4}
\end{equation}%
where $\tilde{v}=\sqrt{v_{\parallel }^{2}+\omega _{0}^{2}\rho _{1}^{2}}$ is
the velocity of the charge image on the cylinder surface. The second
condition in (\ref{peakscond4}) is the Cherenkov condition for the velocity
of the charge image on the cylinder surface and dielectric permittivity of
the cylinder. From (\ref{peakscond3}) it follows that when the Cherenkov
condition for the velocity of the charge and dielectric permittivity of the
surrounding medium is not satisfied, $v\sqrt{\varepsilon _{1}}/c<1$, the
possible strong peaks are located in the angular region defined by the
inequality
\begin{equation}
\left\vert \frac{\tilde{v}}{c}\sqrt{\varepsilon _{1}}\cos \vartheta -\frac{%
v_{\parallel }}{\tilde{v}}\right\vert <\frac{\tilde{v}_{\perp }}{\tilde{v}}%
\sqrt{\frac{\tilde{v}^{2}}{c^{2}}\varepsilon _{0}-1}.  \label{peakregion1}
\end{equation}%
with $\tilde{v}_{\perp }=\omega _{0}\rho _{1}$. If the Cherenkov condition $v%
\sqrt{\varepsilon _{1}}/c>1$ is satisfied, in addition to this inequality we
also need to have the condition
\begin{equation}
\left\vert \frac{v}{c}\sqrt{\varepsilon _{1}}\cos \vartheta -\frac{%
v_{\parallel }}{v}\right\vert >\frac{v_{\perp }}{v}\sqrt{\frac{v^{2}}{c^{2}}%
\varepsilon _{1}-1}.  \label{peakregion2}
\end{equation}

In order to estimate the angular widths for the peaks, we note that near
them the main contribution to the radiation intensity comes from the terms
in the coefficients $D_{m}^{(\pm )}$ containing in the denominators the
function $\alpha _{m}$. Expanding this function near the angle corresponding
to the peak, $\vartheta =\vartheta _{p}$, we see that the angular dependence
of the radiation intensity near the peak has the form
\begin{equation}
\frac{dI_{m}}{d\Omega }\propto \frac{1}{(\vartheta -\vartheta
_{p})^{2}/b_{p}^{2}+1}\left( \frac{dI_{m}}{d\Omega }\right) _{\vartheta
=\vartheta _{p}},  \label{dImnear}
\end{equation}%
where $b_{p}\propto \exp [-2m\zeta (\lambda _{1}\rho _{1}/m)]$. Hence, the
angular widths of the peaks are of the order $\Delta \vartheta \propto \exp
[-2m\zeta (\lambda _{1}\rho _{1}/m)]$. From estimate (\ref{Npeak}) it
follows that at the peaks the angular density of the radiation intensity
exponentially increases with increasing $m$. However, one has to take into
account that in realistic situations the growth of the radiation intensity
is limited by several factors. In particular, the factor which limits the
increase, is the imaginary part for the dielectric permittivity $\varepsilon
_{j}^{\prime \prime }$, $j=0,1$. This leads to additional terms in the
denominator of formula (\ref{dImnear}), proportional to the ratio $%
\varepsilon _{j}^{\prime \prime }/\varepsilon _{j}^{\prime }$, where $%
\varepsilon _{j}^{\prime }$ is the real part of the dielectric permittivity.
As a result for $b_{p}^{2}<\varepsilon _{j}^{\prime \prime }/\varepsilon
_{j}^{\prime }$ the intensity and the width of the peak is determined by
these terms and the saturation of the radiation intensity at the peak takes
place.

\begin{figure}[tbph]
\begin{center}
\begin{tabular}{cc}
\epsfig{figure=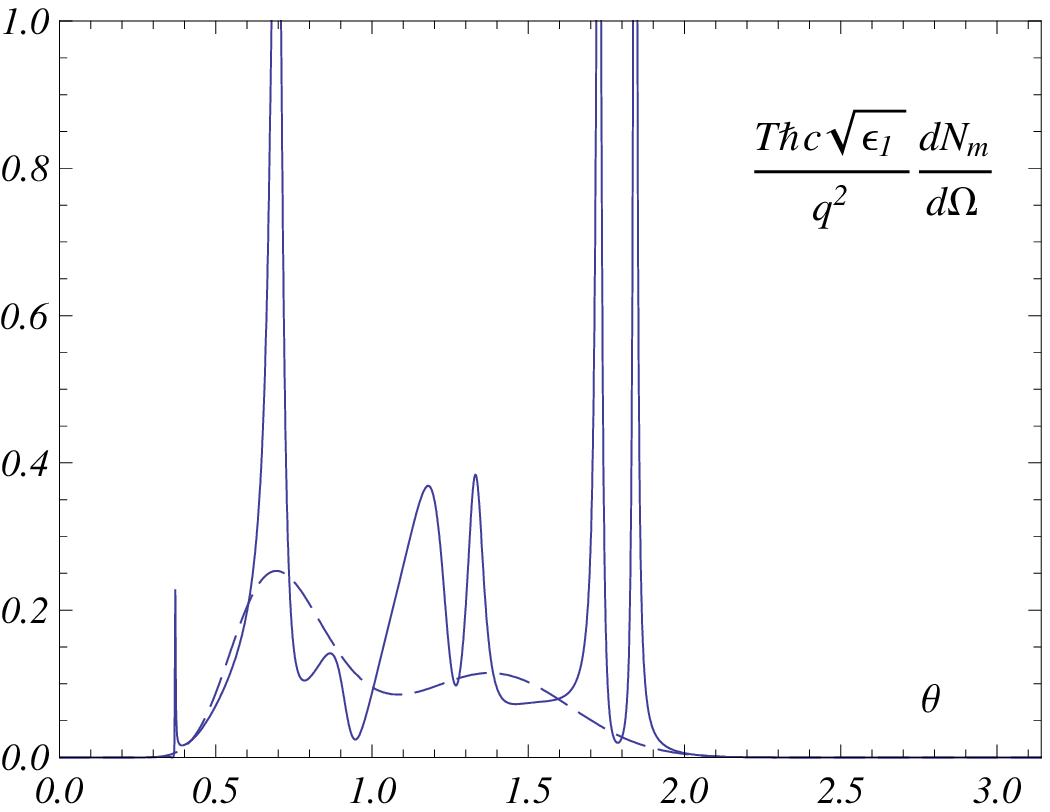,width=6.5cm,height=5.5cm} & \quad %
\epsfig{figure=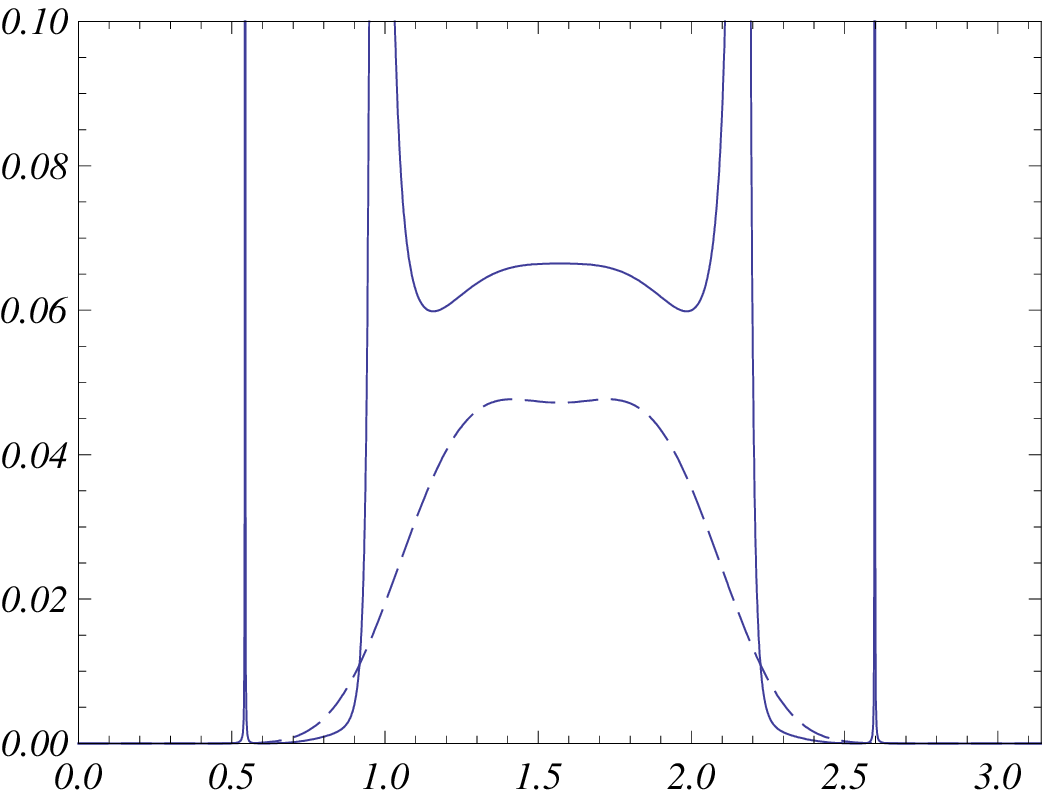,width=6.5cm,height=5.5cm}%
\end{tabular}%
\end{center}
\caption{The dependence of the angular density for the number of radiated
quanta, $(T\protect\sqrt{\protect\varepsilon _1}\hbar c/q^2)dN_m/d\Omega $,
per period $T$ of charge revolution as a function of the angle $%
\protect\vartheta $ for $\protect\beta _{1\perp }=0.9$, $\protect\rho _1/%
\protect\rho _0=0.95$, $m=10$. The left panel is plotted for $\protect\beta %
_{1\parallel }=0.5$ and the right panel is plotted for $\protect\beta %
_{1\parallel }=0$. Full and dashed curves correspond to the cases $\protect%
\varepsilon _0/\protect\varepsilon _1=3$ and $\protect\varepsilon _0/\protect%
\varepsilon _1=1$ respectively. }
\label{fig1}
\end{figure}

By using the general formula (\ref{Int3}), we have carried out numerical
calculations of the radiation intensity at a given harmonic $m$ for various
values of the parameters. As an example, in figure \ref{fig1} we have
plotted the dependence of the angular density for the number of the radiated
quanta
\begin{equation}
\frac{dN_{m}}{d\Omega }=\frac{1}{\hbar \omega _{m}}\frac{dI_{m}}{d\Omega },
\label{dNm}
\end{equation}%
as a function of the angle $0\leqslant \vartheta \leqslant \pi $ for $\beta
_{1\perp }=0.9$, $\rho _{1}/\rho _{0}=0.95$, $m=10$. The full and dashed
curves correspond to the cases $\varepsilon _{0}/\varepsilon _{1}=3$ and $%
\varepsilon _{0}/\varepsilon _{1}=1$ (homogeneous medium) respectively. The
left panel is plotted for the longitudinal component $\beta _{1\parallel
}=0.5$. In order to compare with the radiation intensity in the case of
purely transversal motion, we have plotted on the right panel the
corresponding graphs for $\beta _{1\parallel }=0$. In this case the curves
are symmetric with respect to the rotation plane $\vartheta =\pi /2$. In
both cases strong narrow peaks appear when the dielectric cylinder is
present. At the peaks of the right panel one has $\vartheta \approx 0.5435$,
$(T\sqrt{\varepsilon _{1}}\hbar c/q^{2})dN_{m}/d\Omega \approx 138$, with
the width of the peak $\Delta \vartheta \approx 10^{-4}$, and $\vartheta
\approx 0.973$, $(T\sqrt{\varepsilon _{1}}\hbar c/q^{2})dN_{m}/d\Omega
\approx 12.9$, $\Delta \vartheta \approx 6\cdot 10^{-3}$. For the left peak
on the left panel one has $\vartheta \approx 0.695$, $(T\sqrt{\varepsilon
_{1}}\hbar c/q^{2})dN_{m}/d\Omega \approx 1.53$, $\Delta \vartheta \approx
0.05$; $\vartheta \approx 1.726$, $(T\sqrt{\varepsilon _{1}}\hbar
c/q^{2})dN_{m}/d\Omega \approx 1.67$, $\Delta \vartheta \approx 0.02$; $%
\vartheta \approx 1.842$, $(T\sqrt{\varepsilon _{1}}\hbar
c/q^{2})dN_{m}/d\Omega \approx 2.37$, $\Delta \vartheta \approx 0.01$. These
numerical data are in good agreement with the analytic estimates given
before. From presented graphs it is seen that, for angles away the peaks the
drift essentially amplifies the radiation intensity.

\begin{figure}[tbph]
\begin{center}
\begin{tabular}{cc}
\epsfig{figure=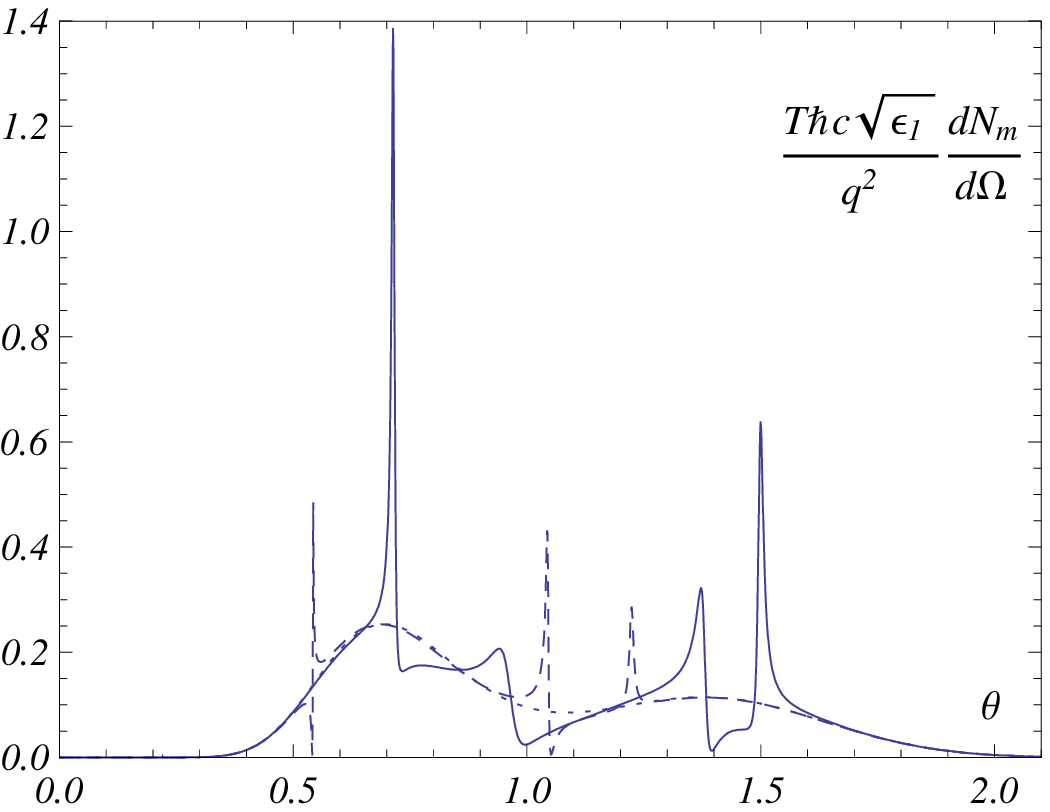,width=6.5cm,height=5.5cm} & \quad %
\epsfig{figure=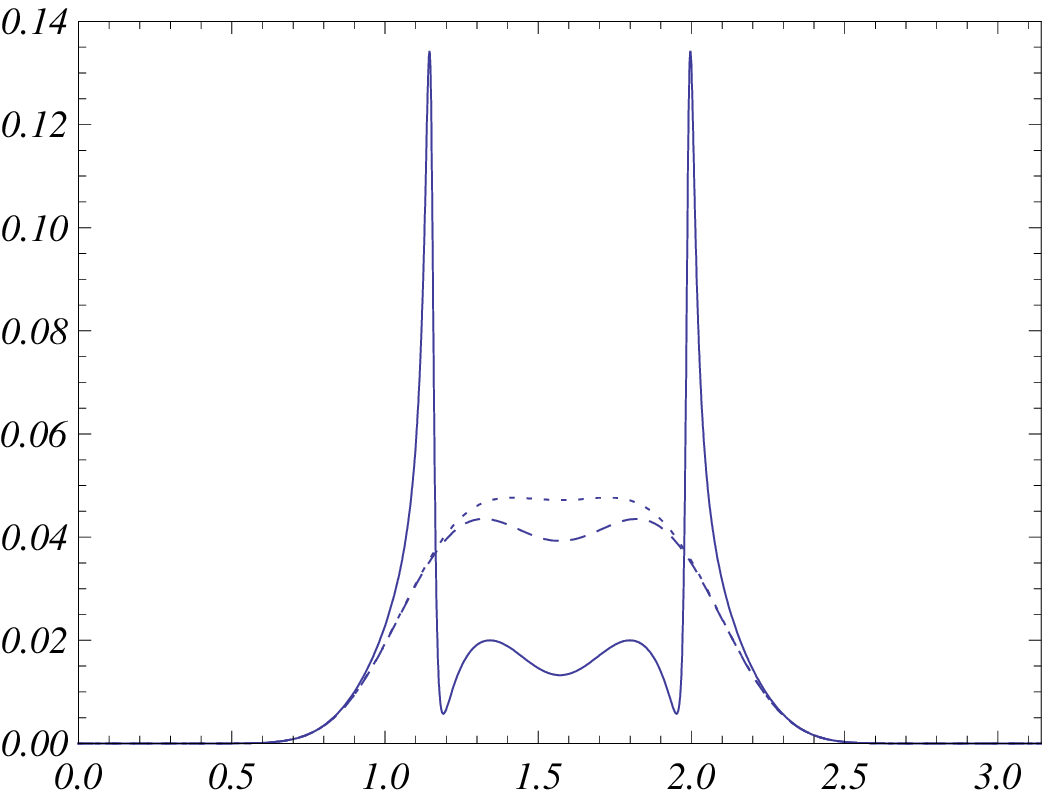,width=6.5cm,height=5.5cm}%
\end{tabular}%
\end{center}
\caption{The dependence of the quantity $(T\protect\sqrt{\protect\varepsilon %
_{1}}\hbar c/q^{2})dN_{m}/d\Omega $ on the angle $\protect\vartheta $ for $%
\protect\beta _{1\perp }=0.9$, $m=10$, $\protect\varepsilon _{0}/\protect%
\varepsilon _{1}=3$. For the left panel $\protect\beta _{1\parallel }=0.5$, $%
\protect\rho _1/\protect\rho _0=0.8$ (full curve), $\protect\rho _1/\protect%
\rho _0=0.7$ (dashed curve). For the right panel $\protect\beta _{1\parallel
}=0$, $\protect\rho _1/\protect\rho _0=0.85$ (full curve), $\protect\rho _1/%
\protect\rho _0=0.8$ (dashed curve). On both panels the dotted lines present
the corresponding quantities for the radiation in a homogeneous medium. }
\label{fig2}
\end{figure}

In order to illustrate the dependence of the radiation intensity on the
ratio $\rho _{1}/\rho _{0}$, in figure \ref{fig2} we have presented the
angular dependence of the number of the radiated quanta for different values
of this ratio. The graphs are plotted for $\beta _{1\perp }=0.9$, $m=10$, $%
\varepsilon _{0}/\varepsilon _{1}=3$. For the left panel $\beta _{1\parallel
}=0.5$. On this panel the full (dashed) curve corresponds to the value $\rho
_{1}/\rho _{0}=0.8$ ($\rho _{1}/\rho _{0}=0.7$). For the right panel $\beta
_{1\parallel }=0$ and the full (dashed) curve corresponds to the value $\rho
_{1}/\rho _{0}=0.85$ ($\rho _{1}/\rho _{0}=0.8$). On both panels the dotted
lines present the corresponding quantities for the radiation in a
homogeneous medium ($\varepsilon _{0}=\varepsilon _{1}$). From this graphs
we see that the influence of the dielectric cylinder is essential only in
the case when the charge trajectory is sufficiently close to the cylinder
surface.

\section{Conclusion}

\label{sec:Conc}

The unique characteristics of synchrotron radiation, such as high intensity,
high collimation, and the wide spectral range, have resulted in its
extensive applications. In the present paper, continuing our previous work
on the influence of a medium on the parameters of the synchrotron radiation,
we have investigated the properties of the radiation from a charged particle
moving along a helical orbit around a dielectric cylinder. In order to
evaluate the corresponding vector potential and the electromagnetic fields
in the exterior medium, we have employed the Green function, previously
investigated in \cite{Grigoryan1995}. These fields are presented in the form
of the Fourier expansion (\ref{vecpot2}) with the Fourier coefficients given
by formulae (\ref{vecpot3c}), (\ref{magneticc}), (\ref{electricc}). On the
base of these formulae, in section \ref{sec:radiation}, we have investigated
the spectral-angular distribution of the radiation propagating in the
exterior medium. In the case when the Cherenkov condition for dielectric
permittivity of the exterior medium and drift velocity of the charge is
satisfied, $\beta _{1\parallel }>1$, the radiation intensity is decomposed
into two terms. The first one, given by formula (\ref{Intm0}), corresponds
to the radiation with continuous spectrum propagating along the Cherenkov
cone of the external medium. The second term in the expression for the total
radiation intensity, given by formula (\ref{Int3a}), presents the
contribution of the harmonics with $m\geqslant 1$. It describes the
radiation, which for a given propagation direction characterized by the
angle $\vartheta $, has a discrete spectrum determined by formula (\ref%
{omegamtet}). This formula describes the normal Doppler effect in
the angular region outside the Cherenkov cone, $\vartheta
>\vartheta _{0}$, and anomalous Doppler effect in the region
inside the cone. In the case $\beta _{1\parallel }<1$, the first
term in (\ref{totintOm}) is absent and the angular distribution of
the radiation intensity at a given harmonic $m$ is given by
formula (\ref{Int3}). In section \ref{sec:Properties} we have
investigated characteristic features of the radiation. In the
non-relativistic limit the general formula for the radiation
intensity reduces to (\ref{Intnonrel}) and the main part of the
radiation energy is emitted on the fundamental harmonic $m=1$. For
a fixed radius of the charge orbit and for small values of the
cylinder radius, for a given harmonic the effects induced by the
presence of the cylinder vanish as $\rho _{1}^{2m}$. As in the
case of the particle motion inside the dielectric cylinder, under
certain conditions on the parameters strong narrow peaks appear in
the angular distribution of the radiation intensity. These peaks
correspond to the values of the angle $\vartheta $ for which the
real part of the function
$\alpha _{m}$ from (\ref{bet1}) vanishes. The corresponding condition (\ref%
{peakscondn}) is obtained from the equation determining the eigenmodes for
the dielectric cylinder replacing the Hankel functions by the Neumann ones.
At the zeros of the real part of $\alpha _{m}$ the imaginary part of this
function is exponentially small for large values $m$ and this leads to
strong angular peaks in the radiation intensity. By using asymptotic
formulae for the cylindrical functions, we have specified the conditions for
the appearance of the peaks and analytically estimated their heights and
widths. In particular, we have shown that the peaks are present only when
dielectric permittivity of the cylinder is greater than the permittivity for
the surrounding medium and the Cherenkov condition is satisfied for the
velocity of charge image on the cylinder surface and the dielectric
permittivity of the cylinder. These features are well confirmed by the
results of the numerical calculations. These results show that the presence
of the cylinder provides a possibility for an essential enhancement of the
radiated power as compared to the radiation in a homogeneous medium.

\section*{Acknowledgement}

The authors are grateful to Professor L.Sh. Grigoryan, S.R. Arzumanyan, H.F.
Khachatryan for stimulating discussions. A.A.S. acknowledges the hospitality
of the Federal University of Paraiba (Jo\~{a}o Pessoa, Brazil). The work has
been supported by Grant No.~0077 from Ministry of Education and Science of
the Republic of Armenia.

\end{document}